# An efficient method based on the evolutionary center algorithm for optimizing chemical-diffusive models for flame acceleration and DDT

Huahua Xiao[a,*], Xu Zhang[a], Mingbin Zhao[a], Congling Shi[b].

[a] *State Key Laboratory of Fire Science, University of Science and Technology of China, Hefei, 230026, China*
[b] *Beijing Key Laboratory of Metro Fire and Passenger Transportation Safety, China Academy of Safety Science and Technology, Beijing, 100012, China*

---

**Abstract**
*This paper presents an efficient method based on the Evolutionary Center Algorithm (ECA) for accurately and efficiently determining the optimal reaction and diffusion parameters for Chemical-Diffusive Models (CDM) to simulate flame acceleration (FA) and deflagration-to-detonation transition (DDT). The proposed method leverages the global search capability of the ECA and the local optimization strength of the Nelder-Mead (NM) algorithm. The hybrid approach (ECA-NM) can efficiently optimize CDM parameters that are capable of accurately reproducing the major properties of combustion waves. The CDMs for premixed flames and detonations of hydrogen in air or oxygen were developed using the present ECA-NM method and validated against canonical tests of combustion waves and previous experiments of FA and DDT. The results show that the major flame and detonation properties calculated using the developed CDMs match those obtained from detailed chemical reaction mechanisms over a wide range of equivalence ratios. The simulated FA and DDT in a channel also agree qualitatively and quantitatively with experiments in terms of complex flame instabilities (e.g., tulip and distorted tulip flames), flame displacement speed, and detonation occurrence. In addition, detailed comparisons to the traditional genetic algorithm demonstrate that the developed ECA-NM method diminishes the global error by nearly four orders of magnitude while reducing the computational cost by two orders of magnitude. This work provides a significantly efficient method for developing chemical-diffusive models that allows quantitative multi-scale simulations of transient flames and detonations in complex scenarios.*

---

*Corresponding author.

## Novelty and significance statement

The efficient and accurate calibration of Chemical-Diffusive Models (CDMs) is critical to their applicability in flame acceleration and deflagration-to-detonation transition simulations. The novelty of this research lies in the development of an Evolutionary Center Algorithm–Nelder Mead (ECA-NM) method that introduces a center-of-mass-guided global search to efficiently identify high quality regions of the CDM parameter space, followed by Nelder-Mead local refinement. This method preserves the fidelity of flame and detonation properties across a broad range of equivalence ratios. In addition, compared with the traditional genetic-algorithm-based calibration strategy, it reduces the calibration cost by two orders of magnitude and the global error by nearly four orders of magnitude. Consequently, this method enables the rapid generation of robust, high-fidelity CDMs, providing an effective basis for quantitative multi-scale simulations of flames, detonations, and transitions between them in complex scenarios.

## 1. Introduction

Efficient and accurate prediction of flame acceleration (FA) and deflagration-to-detonation transition (DDT) in reactive gas mixtures is important for explosion safety[1], detonation-based propulsion systems[2], and supernova in astrophysics[3, 4]. Numerical simulations of FA and DDT rely on the solution of the unsteady, compressible Navier-Stokes equations coupled with suitable chemical reaction models. To provide insightful information about FA and DDT mechanisms, such simulations must resolve the wide range of temporal and spatial scales arising from the interactions of flame, shock, and turbulence[5]. Central to these simulations is the accurate computation of chemical source terms, which directly govern the heat release rate and thereby determine the thermodynamic and hydrodynamic evolution of the flow field [6, 7]. These source terms can be rigorously computed using detailed chemical reaction mechanisms [8], which comprise a set of elementary reactions that describe all of the major and intermediate chemical species and the interactions among them[9, 10]. However, direct application of such detailed mechanisms in multidimensional simulations of FA and DDT imposes an inordinately high computational cost, primarily due to the large number of species and reaction pathways involved. Furthermore, the stiffness arising from widely disparate reaction rates necessitates the use of extremely small-time steps to maintain numerical stability, further compounding the computational burden [11].

In this context, the Chemical-Diffusive Model (CDM) was developed to enhance computational efficiency while maintaining the fidelity of the reactive flow solution [12]. The objective of the CDM is to provide a unified representation for the conversion from reactants to products, energy release, and transport phenomena, governed by a set of adjustable, calibrated parameters. These parameters are calibrated to reproduce the critical flame and detonation properties obtained from experiments, theoretical models, or existing detailed chemical reaction mechanisms. This approach reduces the number of transported species and eliminates the stiffness in the source term calculations. The robustness of the CDM has been extensively validated across a wide range of combustion regimes, including laminar flames[13], turbulent flames[14, 15], shock-flame interactions[6, 16], FA [17, 18] and DDT [19–21].

The predictive accuracy of the CDM largely relies on the calibration of its parameters. In early studies, Kessler et al.[12] employed a graphical approach to determine appropriate input parameters. This procedure entails manually plotting the CDM parameters against the target flame and detonation properties to identify suitable parameters at the intersection points. However, this graphical method is labor intensive, and may give unreliable values for the model parameters. To address this, Kaplan et al.[22] proposed a general automated optimization method based on a combination of Genetic Algorithm (GA) and Nelder-Mead (NM) scheme [23] to enhance the efficiency of parameter search. Specifically, the GA performs a global search through crossover and mutation operations to identify a promising parameter space. The NM scheme is then employed to refine the solution locally. This approach has been proven to be effective in calibrating CDMs for simulating FA and DDT. However, the GA inherently requires a large number of function evaluations to reach the global minimum. As reported in [22], calibrating a stoichiometric methane-air mixture typically requires 3 to 5 hours to achieve convergence under satisfactory error tolerances. To address this computational bottleneck, Lu et al.[24] developed an efficient analytical optimization method that employs asymptotic analysis to derive closed-form expressions for flame and detonation properties, thereby transforming the calibration task into a rapid algebraic root-finding problem.

The purpose of this work is to develop an efficient hybrid method to enhance the efficiency and accuracy of CDM parameter optimization. This method uses a combination of an Evolutionary Center Algorithm (ECA) [25] and the NM optimization scheme to obtain the optimal reaction parameters rapidly and reliably. We describe the algorithmic procedure in detail and then apply it to determine the CDM parameters for hydrogen-air and hydrogen-oxygen mixtures. The fidelity of these parameters is then rigorously validated against canonical tests of combustion waves and experimental data of FA and DDT. Finally, the high efficiency of the proposed ECA-NM method is shown by comparing its performance with that of the traditional GA method.

## 2. Navier-Stokes equations with Chemical-Diffusive model

Simulations of FA and DDT are performed by solving the unsteady, fully compressible, reactive Navier-Stokes equations:

$$\frac{\partial \rho}{\partial t} + \nabla \bullet \left( \rho \vec{U} \right) = 0 , \tag{1}$$

$$\frac{\partial \left( \rho \vec{U} \right)}{\partial t} + \nabla \cdot \left( \rho \vec{U} \vec{U} \right) + \nabla p = \nabla \cdot \hat{\tau} \tag{2}$$

$$\frac{\partial (\rho E)}{\partial t} + \nabla \cdot \left( (\rho E + p) \vec{U} \right) = \nabla \cdot \left( \vec{U} \cdot \hat{\tau} \right) + \nabla \bullet \left( K \nabla T \right) + \rho q \dot{\omega} \tag{3}$$

$$\frac{\partial (\rho Y)}{\partial t} + \nabla \cdot \left( \rho Y \vec{U} \right) = \nabla \cdot \left( \rho D \nabla Y \right) - \rho \dot{\omega} \tag{4}$$

where, $\rho$ is density, $t$ is time, $\vec{U}$ is the velocity vector, $p$ is the pressure, $E$ is the specific total energy, $K$ is the thermal conductivity, $T$ is temperature, $q$ is the chemical energy release, $\dot{\omega}$ is the chemical reaction rate, $Y$ is the mass fraction of the reactant, $D$ is the mass diffusivity. The viscous stress tensor $\hat{\tau}$ is defined using the constitutive equation for Newtonian fluids, which can be written as

$$\hat{\tau} = \rho \nu \left( \left( \nabla \vec{U} \right) + \left( \nabla \vec{U}^{Tr} \right) - \frac{2}{3} \left( \nabla \cdot \vec{U} \right) \mathbf{I} \right) , \tag{5}$$

where $\nu$ is kinematic viscosity, the superscript $Tr$ denotes the transpose operation for a matrix.

The ideal gas law is utilized to describe the relationship between the pressure, volume, and temperature of the gas within the system:

$$p = \frac{\rho R T}{M} , \tag{6}$$

$$E = \frac{p}{(\gamma - 1) \rho} + \frac{1}{2} \left( \vec{U} \cdot \vec{U} \right) , \tag{7}$$

where $M$ is the molecular weight of the reactant, $R$ is the universal gas constant, $\gamma$ is the ratio of specific heats.

As discussed in Section 1, the chemical reaction terms are closed by the CDM, in which the reaction rate is based on a global Arrhenius law to represent the conversion of reactants to products:

$$\dot{\omega} \equiv \frac{dY}{dt} = A\rho Y exp\left(\frac{-E_a}{RT}\right), \quad (8)$$

where $A$ is the pre-exponential factor, and $E_a$ is the activation energy.

The diffusive transport processes are controlled by the temperature-dependent coefficients, including the kinematic viscosity $\nu$, thermal conductivity $K$, and mass diffusivity $D$ in Eqs. (2)-(5). Following a simplified Sutherland's law, thermal conductivity can be calculated as $K = k_0 c_P T^{0.7}$, where $k_0$ is a reference coefficient and $c_p = \gamma R/(\gamma - 1)$ is the constant-pressure heat capacity. The $D$ and $\nu$ also can be calculated if the Lewis number $Le = K/\rho c_p D$ and the Prandtl number $Pr = \rho c_p \nu/K$ are given.

In the CDM [26], the chemical and diffusive processes are governed by a set of adjustable parameters, including the specific heat ratio $\gamma$, the pre-exponential factor $A$, the activation energy $E_a$, the reaction heat release $q$, the mean molecular weight $M$, and the thermal conductivity $k_0$ (other transport relevant coefficients $D_0$, $\nu_0$ can be calculated by Lewis number or Prandtl number). These parameters determine the chemical and transport properties of both reactants and products, thereby controlling the thermal effect and heat release in the combustion process, which in turn govern the global flame velocity and the formation of flame structure. The core strategy of the CDM is to calibrate the above six parameters such that the global properties of the laminar premixed flame and the Zel'dovich-Neumann-Döring (ZND) detonation are reproduced in a manner consistent with detailed chemistry simulations or experiments. At present, the target premixed flame and detonation properties include laminar flame velocity $S_l$, flame thickness $x_{ft}$, adiabatic flame temperature $T_b$, constant volume equilibrium temperature $T_{cv}$, C-J detonation velocity $D_{CJ}$, and the half-reaction width $x_d$.

The properties of one-dimensional (1D) steady-state premixed flame are calculated by solving the ordinary differential equations that describe heat conduction and energy release within the steady-state reaction wave

$$\frac{d}{dx}\left(K\frac{dT}{dx}\right) = \rho\left(U_f c_p \frac{dT}{dx} - q\rho\dot{\omega}\right), \quad (9)$$

$$T(x=-\infty) = T_0 \text{ and } \left.\frac{dT}{dx}\right|_{x=+\infty} = 0, \quad (10)$$

where $U_f = S_l\rho_0/\rho$ is the flow velocity moving with the flame, $\rho_0$ is the density of reactants at initial temperature $T_0$. The flame temperature distribution is obtained by integrating Eq. (9) over the computational domain based on the known initial and boundary conditions. The calculated temperature gradient $dT/dx$ depends on $S_l$. The $S_l$ is determined by a bisection method within the preset range, and the integration procedure is repeated until the $dT/dx = 0$ at $T = T_0 + q/c_p$ At this point, the chemical heat release is entirely used to raise the gas temperature from $T_0$ to $T_b$. Consequently, the $T_b$ and the constant volume equilibrium temperature $T_{cv}$ are given by

$$T_b = T_0 + q/c_p, \quad (11)$$

$$T_{cv} = T_0 + q/c_v. \quad (12)$$

The laminar flame thickness is defined from the temperature gradient as

$$x_{ft} = (T_b - T_0) \Big/ \max\left|\frac{dT}{dx}\right|. \quad (13)$$

The detonation properties are obtained from the ZND model. The analytical expression for the CJ detonation velocity is given by

$$D_{CJ} = c_0\left(\sqrt{1 + \frac{q}{p_0}\frac{\rho_0(\gamma^2-1)}{2\gamma}} + \sqrt{\frac{q}{p_0}\frac{\rho_0(\gamma^2-1)}{2\gamma}}\right), \quad (14)$$

where $c_0$ is the sound speed of the initial condition.

The ZND parameters corresponding to $D_{CJ}$ can be calculated by

$$\frac{p_{ZND}}{p_0} = 2M_{CJ}^2\frac{\gamma}{\gamma+1} - \frac{\gamma-1}{\gamma+1}, \quad (15)$$

$$\frac{\rho_{ZND}}{\rho_0} = \frac{M_{CJ}^2(\gamma+1)}{M_{CJ}^2(\gamma-1)+2}, \quad (16)$$

$$e_{ZND} - e_0 = 0.5(p_{ZND} + p_0)\left(\frac{1}{\rho_0} - \frac{1}{\rho_{ZND}}\right), \quad (17)$$

where $M_{CJ} = D_{CJ}/c_0$, and $e_0 = p_0/(\rho_0(\gamma - 1))$. These parameters serve as initial conditions for integrating the 1D steady equations of reaction zone, governed by

$$\frac{d\rho}{dt} = \frac{q\dot{\omega}\rho(\gamma-1)}{U_s^2 - c^2}, \quad (18)$$

$$\frac{dE}{dt} = \frac{P}{\rho^2}\frac{d\rho}{dt} + q\dot{\omega}, \quad (19)$$

$$\frac{dx}{dt} = U_s, \quad (20)$$

$$U_s = \frac{D_{CJ}\rho_0}{\rho}, \quad (21)$$

where $U_s$ is the flow velocity relative to the shock, $c$ is the sound speed. The reaction zone profile is computed by integrating Eqs. (18) - (21) using a second-order Runge-Kutta method from the von Neumann state to the CJ state. The half-reaction thickness, $x_d$, which is defined as the distance between the leading shock wave and the point where the mass fraction $Y = 0.5$, can be extracted from the reaction zone profile.

The framework described above set of equations Eqs. (9) - (21) establishes a forward mapping from the six input parameters $\gamma$, $A$, $E_a$, $q$, $M$ and $k_0$ to the flame and detonation properties $S_l$, $x_{ft}$, $T_b$, $T_{cv}$, $D_{CJ}$, $x_d$. Consequently, the calibration task is mathematically defined as an inverse problem: identifying the optimal parameter set such that the calculated properties minimize the error relative to the target values obtained from detailed chemical simulations. To solve this, the GA-NM approach in [22] evaluates the fitness of candidate parameter sets by quantifying the discrepancy between the computed and target properties. The subsequent generations are produced by applying crossover and mutation operators to parent solutions selected based on their high fitness.

## 3. Optimization procedure

In this section, we present an efficient hybrid method that uses a combination of an ECA and the NM optimization scheme to rapidly identify the optimal reaction parameters. Specifically, the ECA is employed to perform a rapid global search to identify promising parameter regions, the NM scheme is subsequently applied to locally refine the solution. The theoretical principles of the individual algorithms and the implementation of the coupled ECA-NM strategy are detailed below.

### *3.1 Evolutionary Center Algorithm*

The Evolutionary Centers Algorithm (ECA) is a metaheuristic inspired by the physical definition of the center of mass [27]. It is always used for single-objective optimization problems within D-dimensional continuous search space. The algorithm is characterized by its structural simplicity and a minimal number of control parameters, which significantly simplifies fine-tuning process. Similar to physics, where the center of mass represents the weighted average position of a mass distribution, the center of mass in ECA represents the weighted average position of all elements (solutions) in the current population. In this context, the weights are determined by the objective function value (i.e., fitness value here) of each element. Consequently, elements with higher fitness values exert a greater influence on the position of the center of mass, thereby guiding the evolution toward optimal regions.

The schematic representation of the ECA is shown in Fig. 1. The algorithm works by generating an irregular subset of $K$ elements from the population and then calculating their center of mass to establish a new evolutionary direction for the next generation. The method moves the population towards regions of higher fitness, enabling the algorithm to explore the solution space more efficiently and converge to the global optimal solution faster.

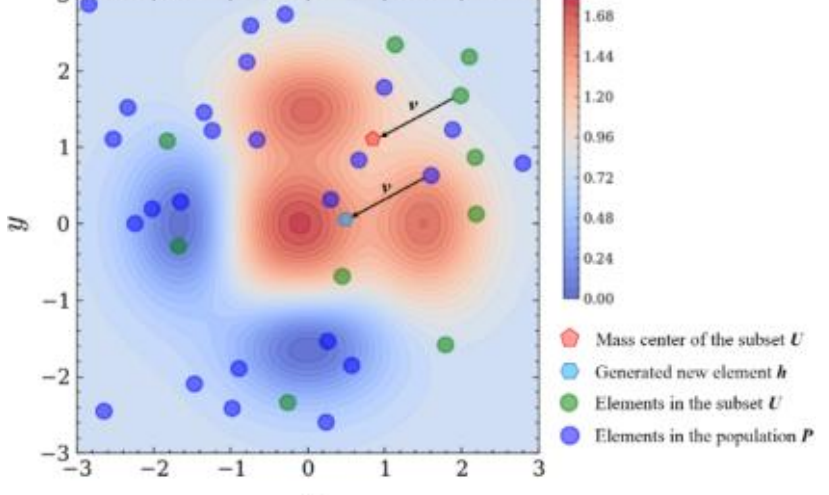

Fig. 1. Schematic representation of the ECA

The algorithm proceeds through the following key steps. First, the control parameters are initialized, including the max step size $\eta_{max}$, the number of elements in the subset $K$, and the number of traits $D$ (which corresponds to the six CDM input parameters). An initial population $P = \{x_1, x_2, \ldots, x_n,\}$ consisting of $N$ elements is then constructed. In each iteration, a subset $U \in P$ of $K$ elements is selected and the center of mass $c_i$ of this subset is calculated as a fitness-weighted centroid

$$c_i = \frac{1}{W}\sum_{u \in U} f(u) \cdot u, \quad W = \sum_{u \in U} f(u) \quad , \tag{22}$$

where $u$ is the elements in the subset $U$, and $f(u)$ is a function used to evaluate the fitness value of each element, defined explicitly in Section 3.3.

Subsequently, based on a randomly chosen solution $u_r \in U$ and the calculated center of mass, a new direction is generated to create a new element $h_i$, following the update rule:

$$h_i = x_i + \eta_i (c_i - u_r) \quad , \tag{23}$$

where $\eta_i$ is a constant selected randomly between $(0, \eta_{max}]$.

The population $P$ is updated with the newly generated elements $h_i$, and this evolutionary process repeats until the end criterion is achieved. The pseudocode of ECA can be found in Fig. 2.

**Algorithm 1: Pseudocode of Evolutionary Centers Algorithm**

**Input**: Max step size ($\eta_{max}$); Number of elements in the subset ($K$); Number of traits of each element ($D$);

**Output**: Best elements in population $P$;

1. Initialize the population $P$ containing $N$ elements $\boldsymbol{x}_i$;
2. **while** the end criterion is not achieved **do**:
3. Create an empty collection $A = \emptyset$;
4. **for** $\boldsymbol{x}_i$ in $P$ **do**:
5. Randomly select $K$ elements from $P$ to generate the subset $U$;
6. Calculate mass center $\mathbf{c}$ of the subset $U$ using Eq. (23);
7. Calculate step size $\eta_i \leftarrow \text{rand}(0, \eta_{max})$;
8. Generate new element $\boldsymbol{h}_i \leftarrow \boldsymbol{x}_i + \eta_i * (\mathbf{c} + \boldsymbol{u}_r)$, where $\boldsymbol{u}_r \in U$ randomly;
9. **if** $(f(\boldsymbol{x}_i) < f(\boldsymbol{h}_i))$ **then**:
10. Append $\boldsymbol{h}_i$ in $A$;
11. **end if**
12. **end for**
13. $P \leftarrow$ best elements in $P \cup A$;
14. **end while**
15. Select best solution in $P$;
16. **end**

Fig. 2. Pseudocode of the ECA

### *3.2 Nelder-Mead algorithm*

The Nelder-Mead (NM) algorithm is a numerical method for solving multi-dimensional unconstrained optimization problems. It is an iterative search algorithm that uses the concept of simplices to find the minimum of a function. The method does not require the information about the derivatives of the objective function, making it particularly useful when dealing with functions that are not differentiable or whose derivatives are complex to compute.

In this method, an initial simplex, composed of $n$+1 points (where $n$ is the dimension of the variable) is constructed. The vertices of the simplex are then transformed by reflection, expansion, contraction (outside and inside), and shrinkage, (as shown in Figure 3) in each iteration to approximate the optimal solution gradually. Each operation requires defining the corresponding coefficients: reflection ($\alpha$), expansion ($\beta$), contraction ($\gamma$), and shrinkage ($\delta$). The details of the NM iteration are given in Figure 4.

The goal of the NM algorithm is to replace the worst point in the simplex with a better point iteratively. However, the algorithm is sensitive to the initial estimate and may fall into local optimal solutions in some cases. Therefore, it is often used in conjunction with other optimization strategies to improve robustness.

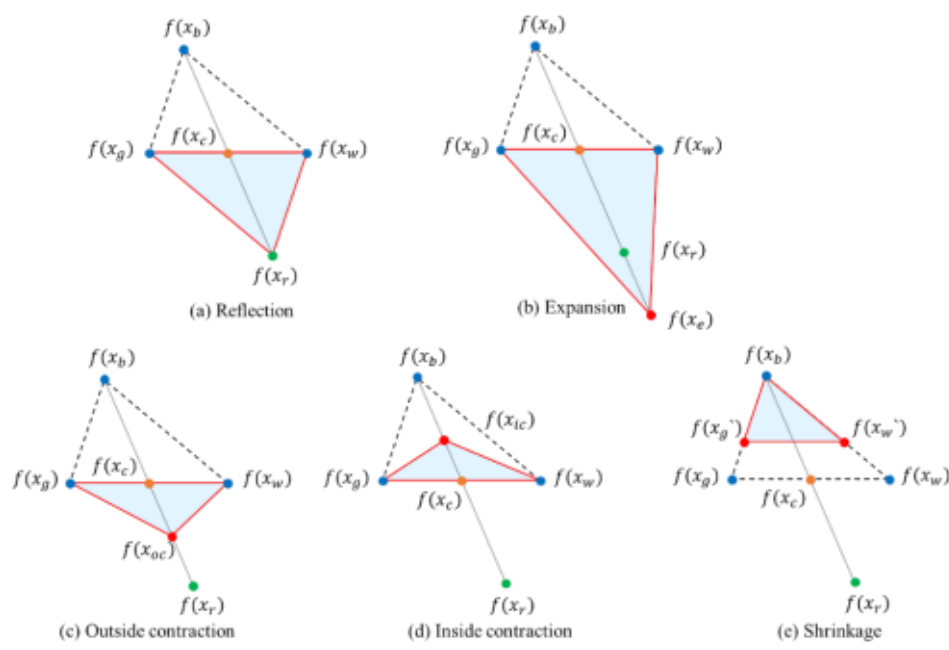

Fig. 3. Schematic illustration of the Nelder-Mead (NM) algorithm simplex operations.

**Algorithm 2: Pseudocode of Nelder-Mead Algorithm**

**Input**: Reflection factor $\alpha$; Expansion factor $\beta$; Contraction factor $\gamma$; Shrink factor $\delta$; Simplex $\boldsymbol{X}$ with $n+1$ vertices $x_i$; Tolerance $\xi$; Function $f(x)$;
**Here use 3 vertices $x_b, x_g, x_w$ as examples to explain the algorithm.**

**Output**: Minimum of function $f(x)$;

1 **while** $(f(\boldsymbol{x}_b) - f(\boldsymbol{x}_g) > \xi)$ **do:**
2 **Ordering:** Order the 3 vertices from $\boldsymbol{X}$ such that $f(\boldsymbol{x}_b) < f(\boldsymbol{x}_g) < f(\boldsymbol{x}_w)$.
3 **Centroid:** Calculate the centroid of 3 vertices $\boldsymbol{x}_c \leftarrow \frac{\boldsymbol{x}_b+\boldsymbol{x}_g+\boldsymbol{x}_w}{3}$.
4 **Reflection:** Calculate the reflection point $\boldsymbol{x}_r = \boldsymbol{x}_c + \alpha(\boldsymbol{x}_c - \boldsymbol{x}_w)$.
5 **if** $(f(\boldsymbol{x}_b) \leq f(\boldsymbol{x}_r) < f(\boldsymbol{x}_g))$ **then:**
6 Accept the reflection point $\boldsymbol{x}_r$ $(\boldsymbol{x}_w \leftarrow \boldsymbol{x}_r)$;
7 **else if** $(f(\boldsymbol{x}_r) < f(\boldsymbol{x}_b))$ **then:**
8 **Expansion:** Calculate the expansion point $\boldsymbol{x}_e = \boldsymbol{x}_c + \beta(\boldsymbol{x}_r - \boldsymbol{x}_c)$.
9 **if** $(f(\boldsymbol{x}_e) < f(\boldsymbol{x}_r))$ **then:**
10 Accept the expansion point $\boldsymbol{x}_e$ $(\boldsymbol{x}_w \leftarrow \boldsymbol{x}_e)$;
11 **else:**
12 Accept the reflection point $\boldsymbol{x}_r$ $(\boldsymbol{x}_w \leftarrow \boldsymbol{x}_r)$;
13 **end if**
14 **else if** $(f(\boldsymbol{x}_g) < f(\boldsymbol{x}_r) < f(\boldsymbol{x}_w))$ **then:**
15 **Outside contraction:** Compute contraction point $\boldsymbol{x}_{oc} = \boldsymbol{x}_c + \gamma(\boldsymbol{x}_r - \boldsymbol{x}_c)$.
16 **if** $(f(\boldsymbol{x}_{oc}) \leq f(\boldsymbol{x}_r))$ **then:**
17 Accept the outside contraction point $\boldsymbol{x}_{oc}$ $(\boldsymbol{x}_w \leftarrow \boldsymbol{x}_{oc})$;
18 **end if**
19 **else**
20 **Inside contraction:** Compute inside contraction point $\boldsymbol{x}_{ic} = \boldsymbol{x}_c - \gamma(\boldsymbol{x}_r - \boldsymbol{x}_c)$.
21 **if** $(f(\boldsymbol{x}_{ic}) \leq f(\boldsymbol{x}_w))$ **then:**
22 Accept the inside contraction point $\boldsymbol{x}_{ic}$ $(\boldsymbol{x}_w \leftarrow \boldsymbol{x}_{ic})$;
23 **end if**
24 **end if**
25 **Shrink:** Replace all points except the best point $x_b$, with $\boldsymbol{x}_i \leftarrow \boldsymbol{x}_b + \delta(\boldsymbol{x}_i - \boldsymbol{x}_b)$.
26 **end while**
27 **end**

Fig. 4. Pseudocode of the Nelder-Mead algorithm.

### *3.3 Implementation of the ECA-NM method*

The hybrid optimization method presented in this work, called ECA–NM, combines the global search capability of the ECA with the local refinement efficiency of the NM algorithm. The overall computational procedure of the ECA-NM algorithm is shown in Figure 5. The optimization procedure begins by establishing the target flame and detonation properties $S_l$, $x_{ft}$, $T_b$, $T_{cv}$, $D_{CJ}$, $x_d$, for a specific fuel-oxidizer mixture and initial conditions. These properties are computed using Cantera [27] and the Shock and Detonation Toolbox [28] with a detailed chemical reaction mechanism.

The objective of the optimization is to minimize the deviation between the CDM-predicted properties and these target values. To achieve this, the optimization process is divided into two stages. First, the ECA is employed to perform a global parameter search, as indicated in the second block of Fig. 5. In this stage, a population is created in which each element xi has six traits corresponding to the CDM input parameters $\gamma$, $A$, $E_a$, $q$, $M$ and $k_0$. For each element, the initial value of traits is randomly selected within the upper and lower bounds. To quantify the quality of each element during the search, a normalized error function and a corresponding fitness function are defined as

$$error = \sqrt{\sum_{i=1}^{ntargets} \left( \frac{\xi_i - \xi_{i,target}}{\xi_i} \right)^2}\ , \qquad (24)$$

$$f = \frac{1}{error + \varepsilon}\ , \qquad (25)$$

where $\xi_i$ is the $i$-th calculated property, the subscript target denotes the corresponding value derived from the detailed mechanism, and $\varepsilon$ is an infinitesimally small value. This fitness definition ensures that parameter sets yielding minimal discrepancies correspond to maximal fitness values.

During the iterative update, new elements with updated trait parameters are generated by targeted mutation operations, as described in Eq. (23). The fitness of each element in the population is then evaluated using Eq. (25) to determine whether the element should be retained in the population. This selection mechanism ensures that traits from highly adapted elements are preserved during the evolutionary process, thereby guiding the population toward the near-optimal solution.

In addition, the test results show that the algorithm often yields multiple optimal solutions. This is primarily because the problem landscape may contain multiple regions with similar fitness values, allowing for various parameter combinations to achieve comparable performance. This is the same phenomenon mentioned in [22]. The most physically plausible set of parameters can be selected by imposing lower and upper bounds on individual reaction parameters. The parameters such as molecular weight ($M$) and the ratio of specific heats ($\gamma$) do not need to perfectly align with detailed thermodynamic values, as long as the key combustion properties are accurately reproduced.

The threshold for ECA stop criteria can be $10^{-1} \sim 10^{-3}$ based on the selection of iteration bounds and reaction mechanism. Once the best solution from the ECA satisfies the error requirements, the optimization procedure transitions to the second stage for local refinement, as indicated in the third block of Fig. 5. In this stage, the NM algorithm is initialized using the final output from the ECA. A simplex with seven vertices is constructed to explore the vicinity of the ECA solution, where one vertex is the ECA solution, and the remaining six vertices are initialized by

$$\left(v_1^{NM}, v_2^{NM} \ldots v_i^{NM}\right) = \left(v_1^{ECA}, v_2^{ECA} \ldots v_i^{ECA}\right) + \left(0, \ldots, v_i^{ECA} \cdot b + a\right), \quad (26)$$

where, $v_i^{ECA}$ epresents the $i$-th component of the CDM input parameters obtained from the ECA output and $v_i^{ECA}$ represents the $i$-th parameter of a vertex need to be initialized. The constants $a$ and $b$ are predefined that control the initialization of the additional vertices. The first part of the equation retains the ECA solution, while the second part introduces variations to explore the solution space around the ECA result. Here, the $a$ = 0.05 and $b$ = 0.00025. The coefficients of reflection ($\alpha$), expansion ($\beta$), contraction ($\xi$), and shrinkage ($\delta$) are calculated by an adaptive scheme based on the dimension of simplex [29], as depicted in Eq. (27). The NM algorithm iteratively replaces the worst point with the better ones and gradually narrows the search range until error reaches the threshold 10-4~10-5 based on the selection of iteration bounds and reaction mechanism.

$$\alpha = 1, \beta = 1 + \frac{2}{n}, \xi = 0.75 - \frac{1}{2n}, \delta = 1 - \frac{1}{n}, \quad (27)$$

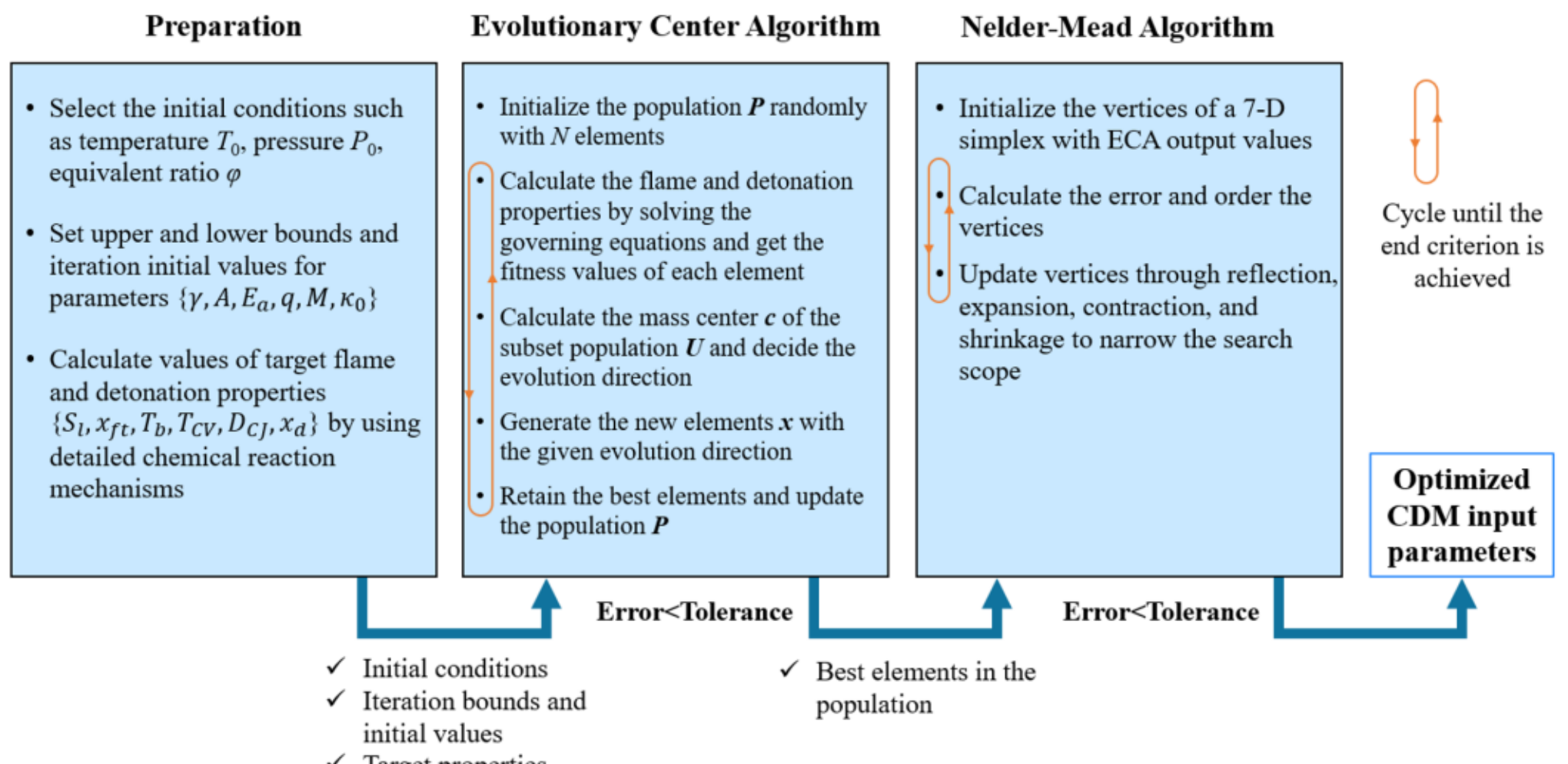

Fig. 5. Overall procedure of the ECA-NM algorithm.

## 4. Results and Validation

In this section, the CDMs for hydrogen-air ($H_2$-air) and hydrogen-oxygen ($H_2$-$O_2$) mixtures are developed using the proposed ECA-NM method and validated against canonical tests of combustion waves and experiments of FA and DDT. Furthermore, the computational efficiency and accuracy of the ECA-NM method are quantitatively compared with those of the GA-NM method.

### *4.1 Calibration results for $H_2$-air and $H_2$-$O_2$ mixtures*

The CDM parameters were calibrated for stoichiometric $H_2$-air and $H_2$-$O_2$ mixtures under initial conditions ($T_0$ = 298 K, $P_0$ = 101325 Pa). The target flame and detonation properties were calculated using the detailed chemical reaction mechanism proposed by Burke et al.[30]. Specifically, the laminar flame properties were simulated using the free-propagating flame in Cantera, and the detonation-related parameters $x_d$ and $D_{CJ}$ were solved using the Shock and Detonation ToolBox. The search space boundaries for each CDM parameter are defined in Table 1. The optimized parameters obtained from the ECA-NM method are shown in Table 2. The corresponding flame and detonation parameters can be obtained by substituting these parameters into the governing equations Eqs.(9)-(21). and the comparison results with the detailed mechanism are shown in Table 3. The global errors, as described by Eq. (24), are 9.7508e$^{-6}$ and 9.6211e$^{-4}$ for the $H_2$-air and $H_2$-$O_2$ mixtures, respectively. These low error values indicate that the reaction parameters calibrated by the ECA-NM method successfully reproduce the intended flame and detonation properties.

Table 1
Search space boundaries for the CDM parameters.

| CDM parameters | Upper and lower bound | |
| --- | --- | --- |
| | Stoichiometric $H_2$-air | Stoichiometric $H_2$-$O_2$ |
| $\gamma$ | 1.0< $\gamma$ <2.0 | 1.0< $\gamma$ <1.2 |
| $E_a$ | 20< $E_a$ <100 | 20< $E_a$ <100 |
| $q$ | 20< $q$ <100 | 20< $q$ <100 |
| $A$ (cm$^3$/g-s) | $10^{10}$< $A$ <$10^{18}$ | $10^{10}$< $A$ <$10^{18}$ |
| $k_0$ (g/s-cm-K$^{0.7}$) | $10^{-6}$< $k_0$ <$10^{-4}$ | $10^{-6}$< $k_0$ <$10^{-4}$ |
| $M$ (g/mol) | 10< $M$ <30 | 10< $M$ <20 |

Table 2
Optimized CDM parameters for stoichiometric $H_2$-air and $H_2$-$O_2$ mixtures ($T_0$ = 298 K, $P_0$ = 101325 Pa).

| CDM parameters | Optimal value | |
|---|---|---|
| | Stoichiometric $H_2$-air | Stoichiometric $H_2$-$O_2$ |
| $\gamma$ | 1.180 | 1.155 |
| $E_a$ | 53.641 $RT_0$ | 54.117 $RT_0$ |
| $q$ | 45.872 $RT_0/M$ | 69.419 $RT_0/M$ |
| $A$ (cm$^3$/g-s) | $8.172\times10^{12}$ | $2.466\times10^{13}$ |
| $k_0$ (g/s-cm-K$^{0.7}$) | $2.557\times10^{-5}$ | $3.720\times10^{-5}$ |
| $M$ (g/mol) | 24.343 | 14.939 |

Table 3
Comparison of flame and detonation properties.

| Properties | $H_2$-air | | $H_2$-$O_2$ | |
|---|---|---|---|---|
| | Burke [30] | ECA-NM | Burke [30] | ECA-NM |
| $T_b$ (K) | 2386.7057 | 2386.7068 | 3074.3830 | 3072.3268 |
| $S_l$ (m/s) | 2.3119 | 2.3119 | 9.7330 | 9.7331 |
| $x_{ft}$ (mm) | 0.3453 | 0.3453 | 0.2376 | 0.2375 |
| $T_{cv}$ (K) | 2763.4163 | 2763.4124 | 3499.5849 | 3502.0193 |
| $D_{CJ}$ (m/s) | 1977.1086 | 1977.0012 | 2839.6298 | 2839.6445 |
| $x_d$ (mm) | 0.1972 | 0.1972 | 0.04635 | 0.04636 |
| Error | $9.7508\times10^{-6}$ | | $9.6211\times10^{-4}$ | |

### *4.2 Validation of 1D flame and detonation structures*

To validate the optimized CDM parameters within a CFD framework, 1D transient simulations were conducted using the reactive Navier-Stokes equations described in Section 2. The numerical domain spans a length of 50 mm, initialized with a stoichiometric $H_2$-air mixture and a 1 mm high-temperature ignition zone at the left boundary. The domain is discretized into 5000 computational cells, yielding a grid resolution of $\Delta x$ = 0.01 mm. This resolution ensures approximately 30 computational cells across the flame thickness, which is sufficient to resolve the flame structure. The convection fluxes are reconstructed using the fifth-order WENO scheme combined with the HLLC Riemann solver, and the diffusion fluxes are discretized by the fourth-order central difference scheme. The time integration is advanced using a third-order Runge–Kutta algorithm. The left boundary of the computed domain is the non-reflective outlet boundary condition, and the right side is the reflection boundary condition.

The resulting 1D flame and detonation wave structures are shown in Fig. 6. In Fig. 6 (a), the temperature and species profiles obtained by solving the transient CFD simulation (symbols) are compared against the steady state solutions (solid lines) derived from Eqs. (9) - (21). The calculated flame profile by CDM and numerical method exhibit excellent agreement with the results obtained from the steady state equations, indicating that the calibrated parameters accurately reproduce the theoretical laminar flame structure. Fig. 6 (b) presents the ZND detonation structure. The profile obtained by integrating the steady state ZND equations with the optimized CDM parameters to the CJ point. The calculated half-reaction width of the 1D detonation wave, $x_d$, exhibits excellent agreement with the detailed mechanism benchmarks.

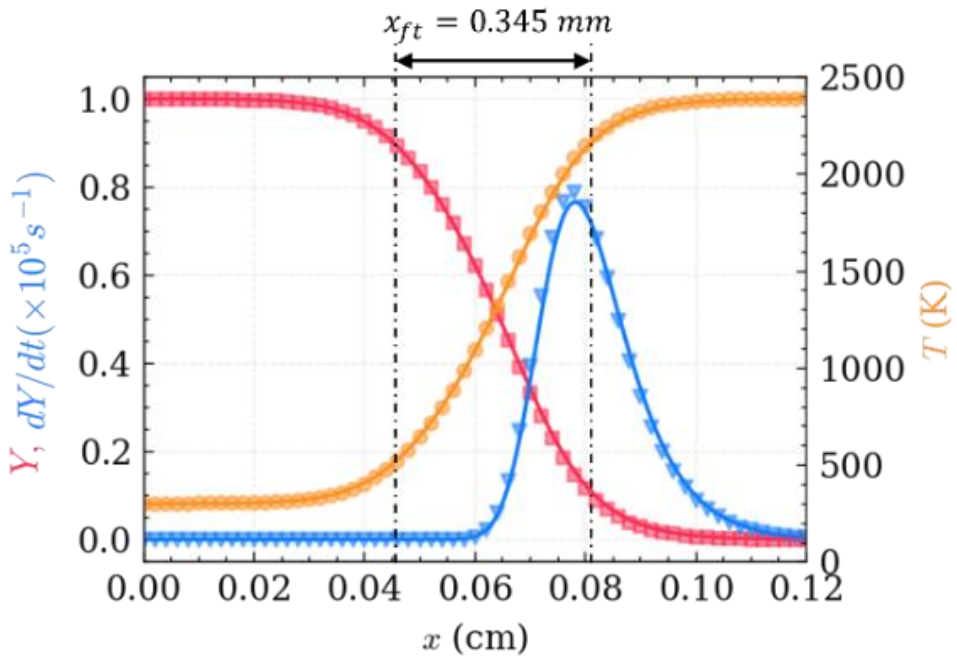

(a). 1D Laminar flame structure

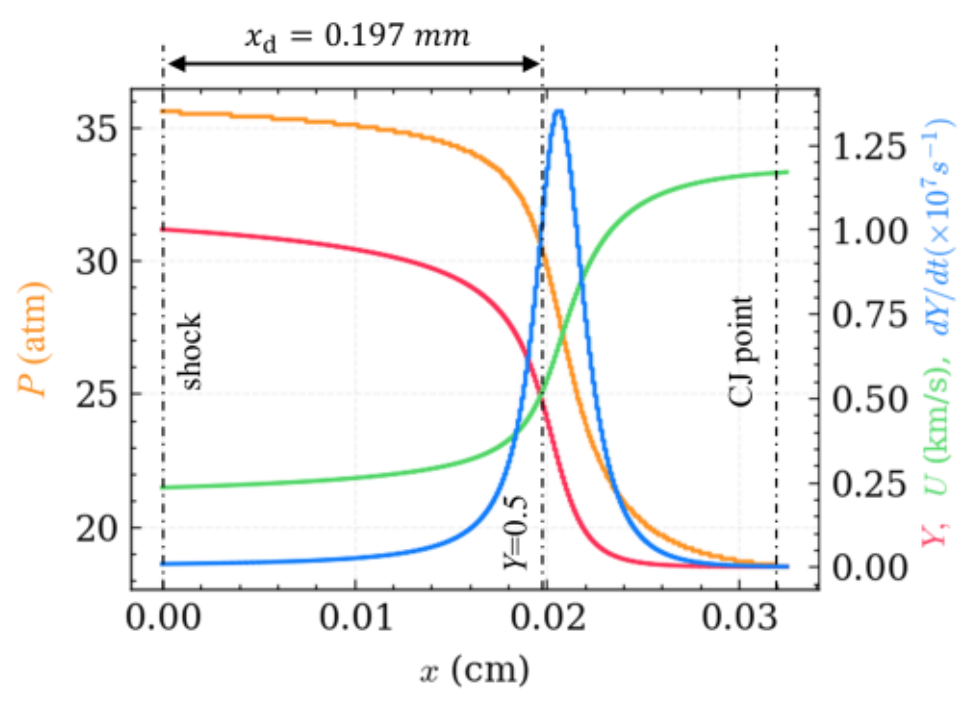

(b) . 1D ZND detonation wave structure

Fig. 6. Computed profiles of the laminar flame and detonation wave structures.

To evaluate the universality and robustness of the ECA-NM method, the CDM parameters were optimized across a broad range of $H_2$-air equivalence ratios($\Phi$ = 0.5~2.5), covering fuel-lean, stoichiometric, and fuel-rich conditions. The laminar flame speeds and adiabatic flame temperatures calculated using the optimized CDM are compared with those obtained from the Burke detailed mechanism under various equivalence ratios, as shown in Fig. 7. The specific optimized CDM parameters and the corresponding characteristic parameters for each equivalence ratio are provided in the Supplementary Material. The comparison shows that as the equivalence ratio varies from lean to rich, both the calculated laminar flame speeds and adiabatic flame temperatures exhibit a trend of initially increasing and then decreasing. Crucially, the flame properties calculated using the CDM show good agreement with those obtained from the detailed chemical mechanism. The relative errors for all tested conditions are maintained within 1%, demonstrating that the ECA-NM method consistently yields high-fidelity parameters across a wide range of equivalence ratios.

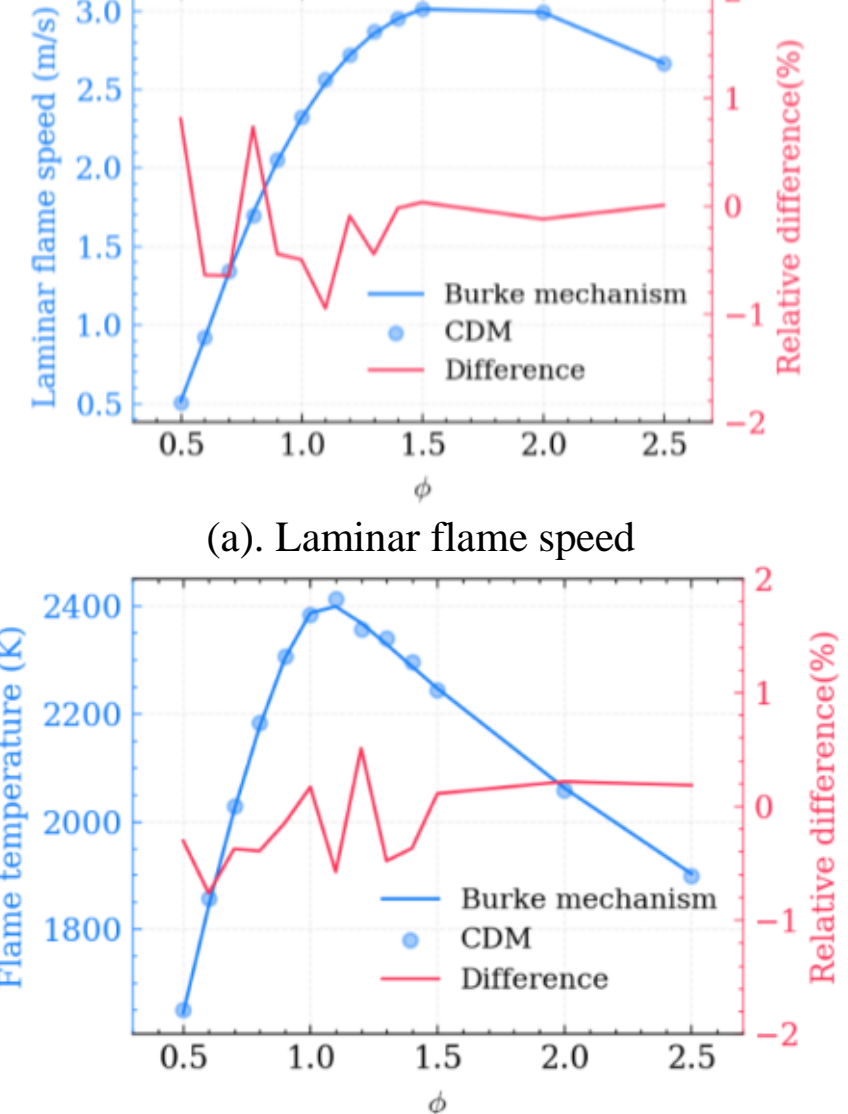

(b). Adiabatic flame temperature

Fig. 7. The laminar flame speed and adiabatic flame temperature as a function of equivalence ratio

### 4.3 *Validation of 2D flame structure and dynamics*

In this section, the CDM parameters optimized from the ECA-NM method are applied to simulate two distinct 2D combustion phenomena. These simulations will evaluate how well the optimized CDM parameters reproduce the complex dynamics of flame evolution and DDT.

#### *4.3.1 Tulip flame and distorted tulip flame*

A tulip flame (TF) is characterized by a concave structure at the flame front, which under specific conditions evolves into a distorted tulip flame (DTF) featuring secondary cusps and intricate wrinkles along the flame front [31, 32]. These structures arise from the complex coupling between chemical kinetics and flow instabilities. Consequently, the simulation of TF and DTF evolution provides a rigorous test to verify the fidelity of the optimized CDM parameters in capturing dynamic flame behaviors.

The computation domain replicates the experimental configuration [13,32], consisting of a closed rectangular channel with a length of $l$ = 53 cm and a width of $d$ = 8.2 cm, as shown in Fig. 8. Adiabatic, no-slip boundary conditions are applied to all wall surfaces. The channel is initialized with a stoichiometric $H_2$-air mixture, where $T_0$ = 298 K and $P_0$ = 101325 Pa. The flame is ignited by placing a semi-circular pocket of hot, burned gas at 101325 Pa and 2140 K. The ignition region has a radius of 0.5 mm and is centered at the midpoint of the left wall. Numerical simulations were performed by solving the governing equations established in Section 2. The Adaptive Mesh Refinement (AMR) [33] is used to dynamically refine the regions of critical flow features. The base grid size used in the current calculations was $dx_{max}$ = 1/80 cm, with the minimum grid size being $dx_{min}$ = 1/640 cm. This resolution ensures approximately 22 cells within the flame front, which is sufficient for resolving the flame structure. The spatial and temporal discretization schemes are consistent with those detailed in Section 4.2.

Fig. 9 illustrates the temporal evolution of the simulated flame front compared with experimental observations [13]. The numerical results successfully reproduce the key morphological transitions of the flame. Specifically, at $t$ = 5.667 ms, the flame front exhibits a distinct flattening behavior, marking the precursor to the flame inversion. Subsequently, the flame front collapses inward to establish the classical concave tulip structure. At $t$ = 7.333 ms, the tulip flame front begins to distort, leading to the formation of the DTF. The numerical simulations accurately capture these complex morphological changes and their timing, showing excellent agreement with the experimental observations.

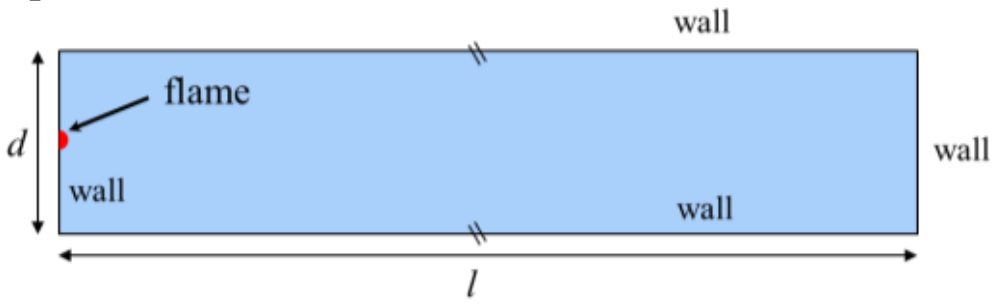

Fig. 8. Schematic diagram of computation domain.

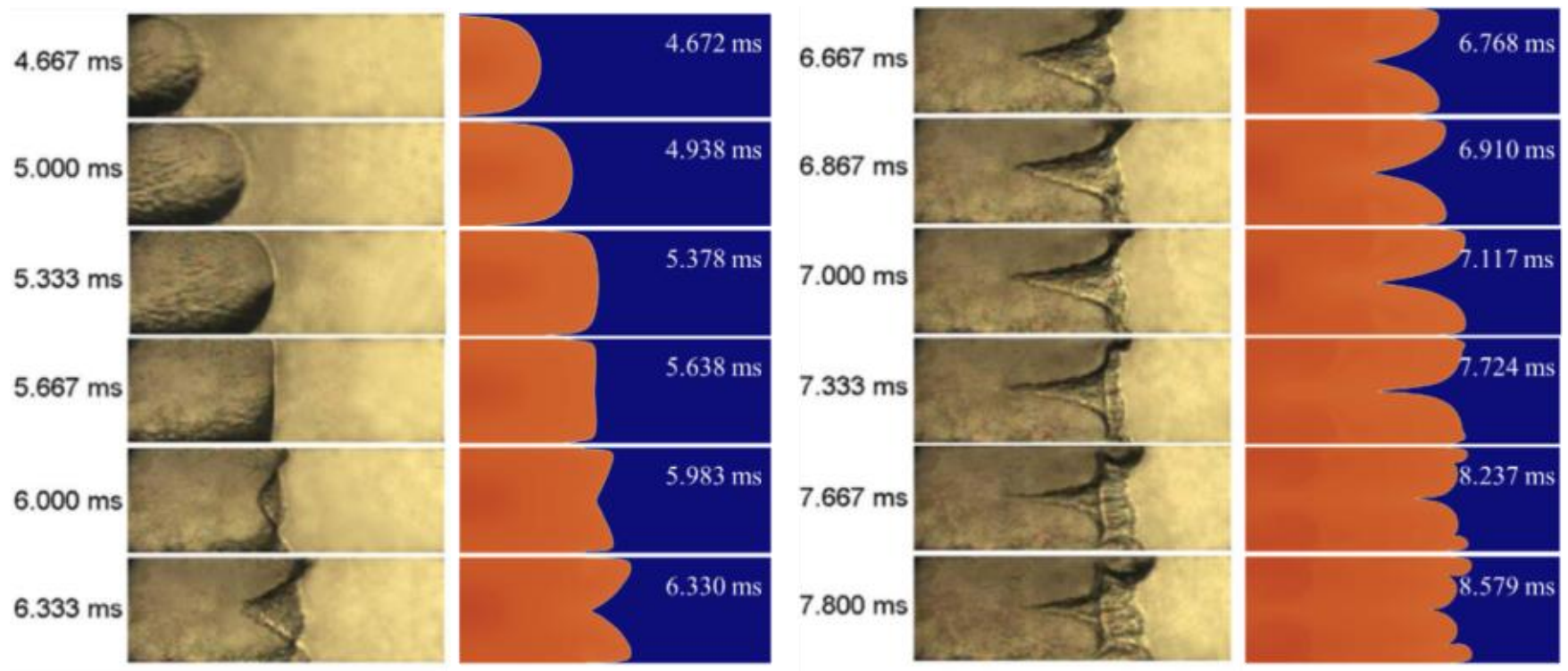

Fig. 9. Comparison of premixed $H_2$-air flame morphology in the experiment [13] and simulation.

*4.3.2 FA and DDT in the obstructed channels*

The simulation of FA and DDT in an obstructed channel serves as a rigorous benchmark to reproduce the intense coupling between shock waves, turbulence, and chemical kinetics. The computation domain, illustrated in Fig. 10, consists of a rectangular channel with dimensions $d \times l$, with continuous arrays of triangular obstacles arranged along both the upper and lower walls. The channel is initialized with a stoichiometric $H_2$-$O_2$ mixture at $T_0$ = 298 K, $P_0$ = 101325 Pa. Adiabatic, no-slip boundary conditions are imposed on the left, upper, and lower walls, as well as on the obstacle surfaces, while a non-reflective boundary condition is applied at the right outlet using NSCBC (Navier-Stokes Characteristic Boundary Condition) approach [34]. The obstacles are modelled using an immersed boundary method (IBM) [35]. The numerical schemes and AMR employed here are identical to those detailed in Section 4.3.1.

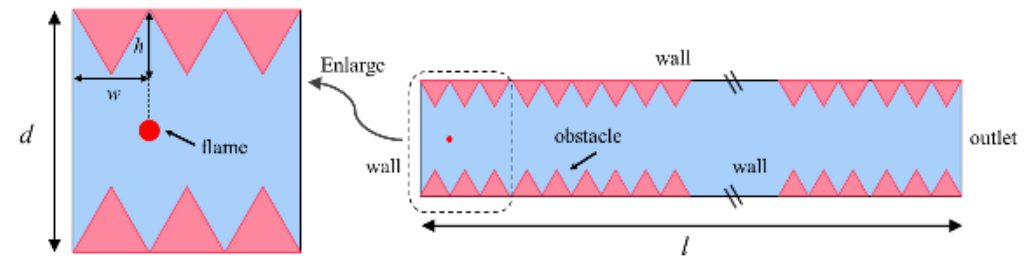

Fig. 10. Schematic diagram of computation domain.
$d$= 2 cm, $l$=30 cm, $w$=0.58 cm, $h$=0.5 cm.

Fig. 11 compares the experimental and numerical schlieren images of the flame front at approximately the same position. Qualitatively, the numerical simulations successfully reproduce the entire sequence of FA and DDT observed in the experiments. In the early stages (e.g., at 0.1282 ms in Figure 11 (b)), the flame expands outward, generating a series of vortices in the gaps between obstacles ahead of the flame. Subsequently, as the flame accelerates down the channel, compression waves coalesce ahead of the flame front, intensifying into a leading shock. At approximately 0.2 ms, the leading shock wave triggers local explosions near the obstacles, ultimately leading to DDT.

Figure 12 compares the flame front speed as a function of position (a) and the flame front position as a function of time (b) between the experimental and numerical results. The results show that the simulated reaction front velocity and the run-up distance to the DDT exhibit excellent agreement with the experimental measurements. The velocity profiles indicate that after DDT, the detonation propagates at an approximately constant speed of DCJ = 2839.61 m/s, consistent with the theoretical value. These computational results demonstrate that the calibrated CDM parameters can accurately reproduce the key dynamics of FA and DDT.

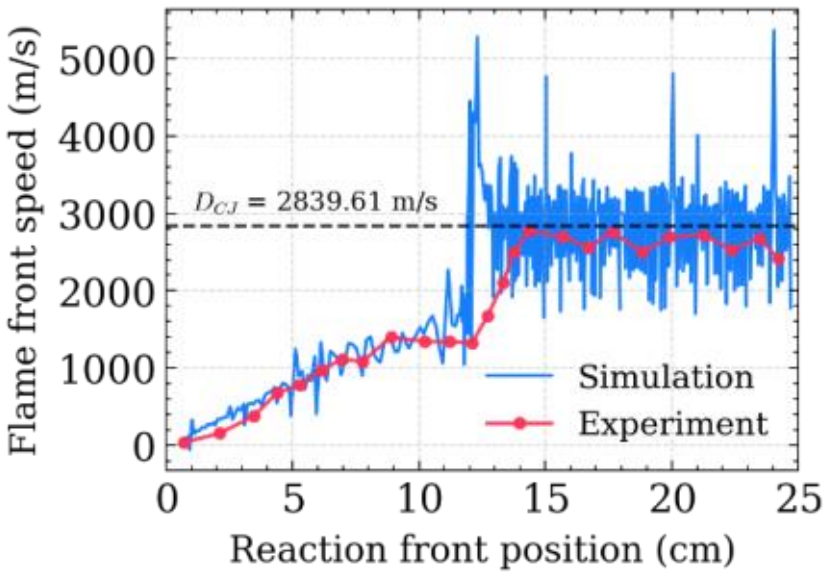

(a) Flame front speed as a function of position

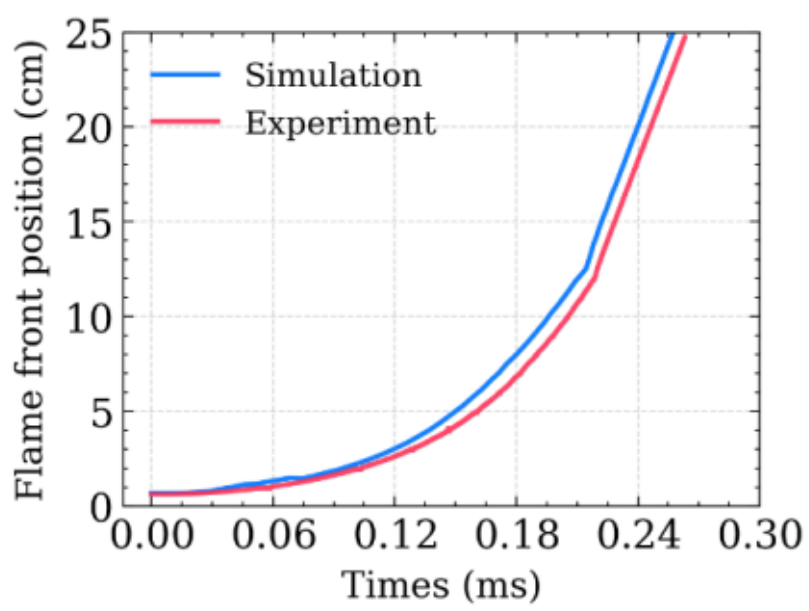

(b) Flame front position as a function of time.

Fig. 12. Comparison between experimental and numerical results of the detonation wave.

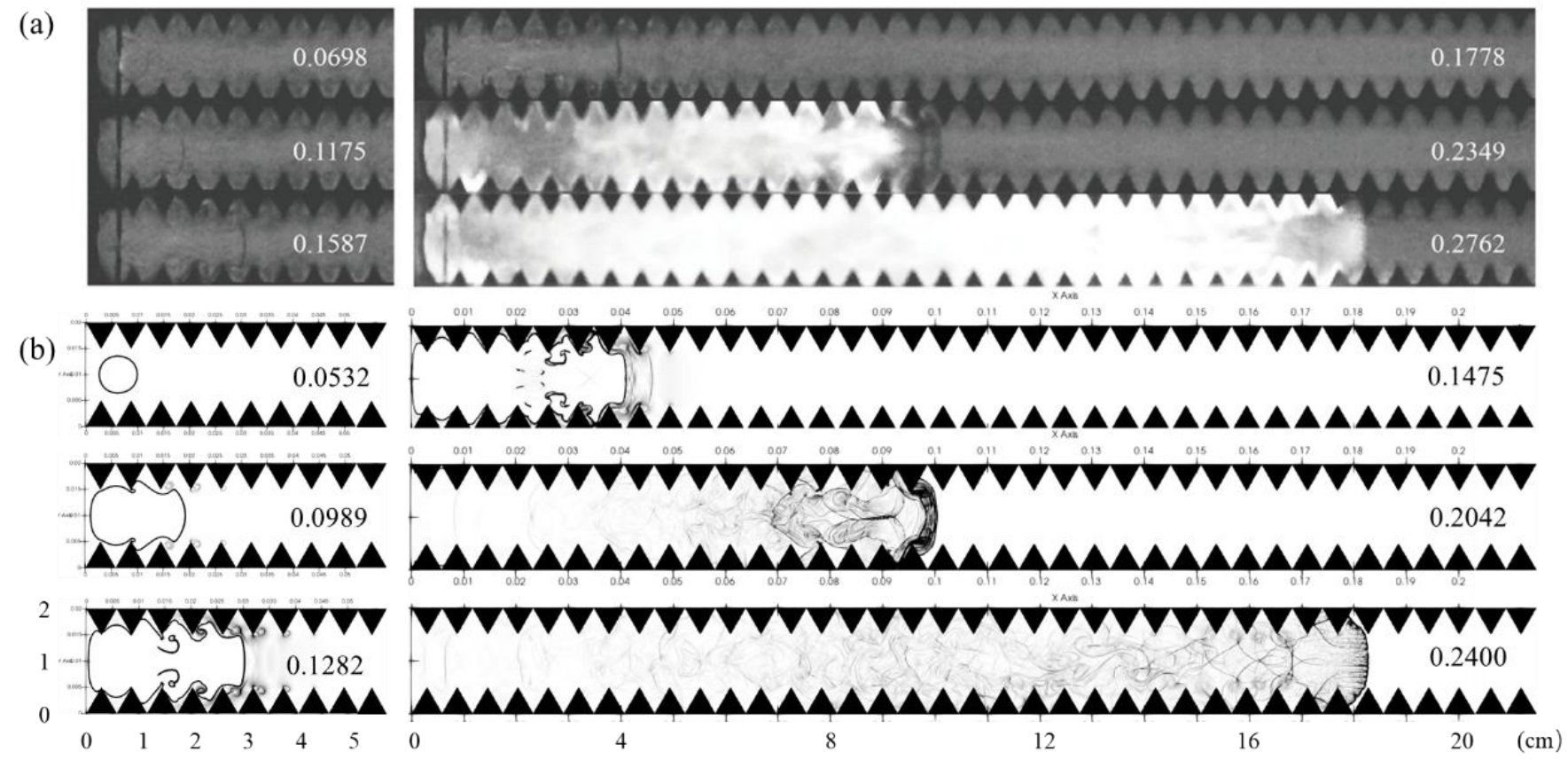

Fig. 11. Comparison of experimental and numerical schlieren images of flame acceleration and DDT in the obstacle channel.
(a) Experimental observations reproduced from [36]; (b) Numerical results obtained using the optimized CDM.

### *4.4 Comparison between GA-NM and ECA-NM method*

The convergence speed and computational accuracy of the proposed ECA-NM method were evaluated against the traditional GA-NM method using the calibration of the stoichiometric $H_2$-air mixture reaction parameters. To ensure a fair comparison based on established and validated configurations, the population parameters for the GA and ECA methods were adopted from the reference in [22] and [37], respectively. The search boundaries for all CDM parameters are identical for both methods, as shown in Table 1. The stopping criteria for the global search algorithms (ECA and GA) were set to a computational error tolerance of $10^{-3}$ or a maximum of 1000 iterations. For the NM algorithm, the criteria were set to an error tolerance of $10^{-5}$ or a maximum of 10000 iterations. These criteria were selected to strike a balance between solution precision and efficiency.

Table 4
Optimal CDM parameters calculated from ECA-NM and GA-NM method for stoichiometric $H_2$-air mixture.

| CDM parameters | Optimal value | |
|---|---|---|
| | ECA-NM | GA-NM |
| $\gamma$ | 1.1804 | 1.1829 |
| $E_a$ | 53.641 $RT_0$ | 55.702 $RT_0$ |
| $q$ | 45.872 $RT_0/M$ | 45.292 $RT_0/M$ |
| $A$ (cm$^3$/g-s) | $8.172\times10^{12}$ | $1.283\times10^{13}$ |
| $k_0$ (g/s-cm-K$^{0.7}$) | $2.557\times10^{-5}$ | $2.404\times10^{-5}$ |
| $M$ (g/mol) | 24.343 | 22.689 |

Fig. 13 illustrates the changes in error with the number of iterations for both the ECA-NM and the GA-NM. The ECA demonstrates superior convergence properties, achieving the error tolerance of $10^{-3}$ only 203 iterations, as shown in Fig. 13(a). In contrast, while the GA initially finds a direction with fast convergence, it appears to get trapped in a local minimum, causing the error to fluctuate around 0.14 without further reduction until the end of the computation, as shown in Fig. 13(c). The NM algorithm was employed to further optimize the results obtained from these search algorithms. The impact of the initial guess on the subsequent NM optimization is evident. The ECA-optimized solution, being close to the optimal, allows the NM algorithm to converge rapidly to the stricter tolerance ($10^{-5}$). Conversely, the GA-provided solution contains a significant residual error. The NM algorithm fails to locate a solution meeting the stopping criteria within the neighborhood of the GA result, terminating only after reaching the maximum iteration count.

Table 5 presents a comprehensive comparison of the calibrated parameters, global errors, and computational costs between ECA-NM and GA-NM. Computational results obtained on an Intel Core i5-10400F processor (@ 2.90GHz) reveal a substantial efficiency gap: the GA-NM required 21752.30 s to complete the calibration, whereas the ECA-NM finished in just 113.32 s, representing a speedup of approximately two orders of magnitude. In addition to this significant efficiency gain, the ECA-NM method demonstrates superior accuracy. The global error for the ECA-NM solution is $9.7508\times10^{-6}$, which is nearly four orders of magnitude lower than that of the GA-NM ($5.3725\times10^{-2}$). Consequently, the flame and detonation properties calculated by ECA-NM match the target values with near-exact precision, whereas the GA-NM results exhibit some deviations.

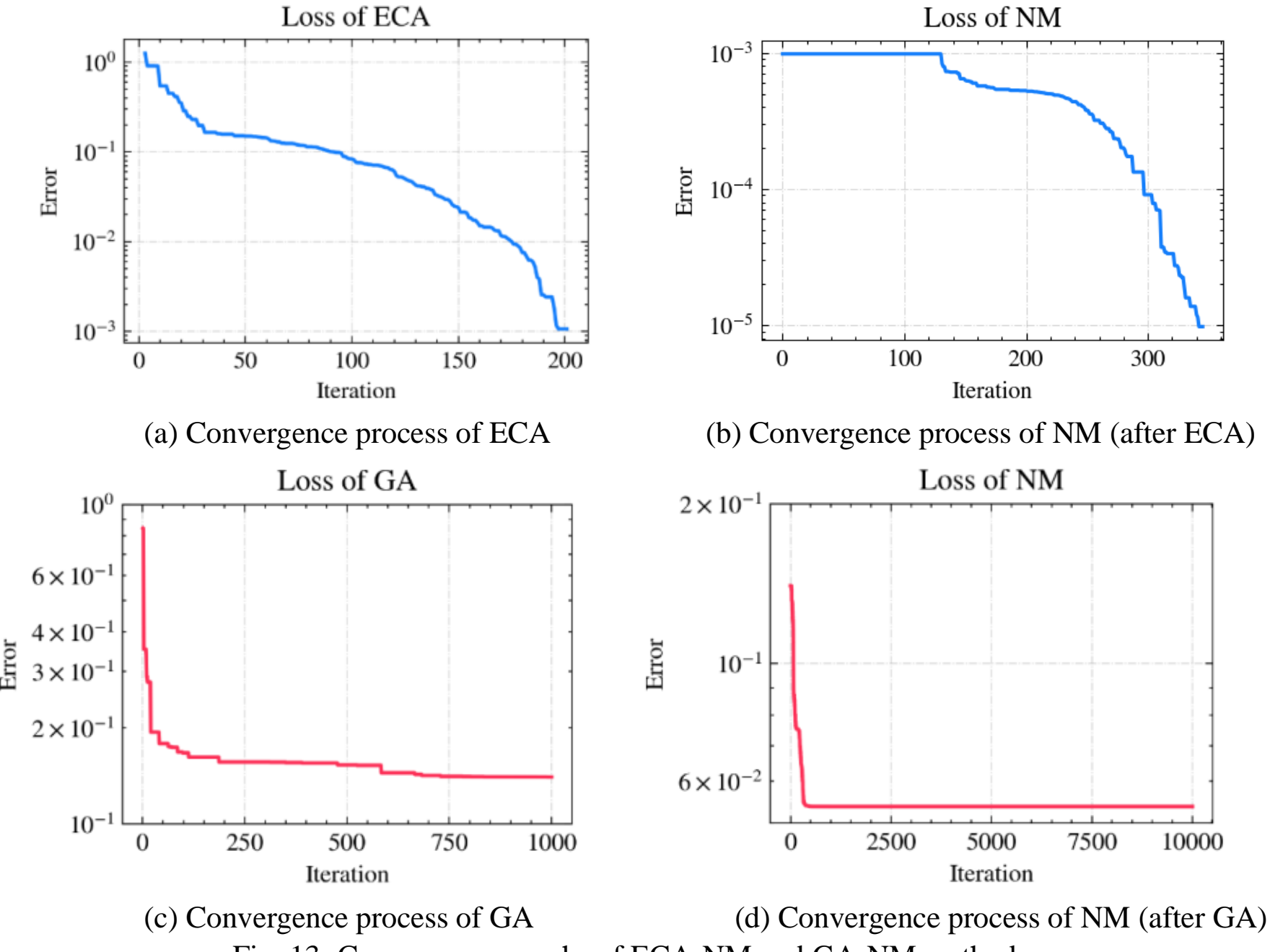

(a) Convergence process of ECA
(b) Convergence process of NM (after ECA)
(c) Convergence process of GA
(d) Convergence process of NM (after GA)

Fig. 13. Convergence graphs of ECA-NM and GA-NM method.

Table 5
Comparison of calibrated reaction parameters, global errors, and computational costs between the ECA-NM and GA-NM.

| Properties | Burke [30] | ECA-NM | GA-NM |
|---|---|---|---|
| $T_b$ (K) | 2386.7057 | 2386.7068 | 2385.7605 |
| $S_l$ (m/s) | 2.3119 | 2.3119 | 2.3745 |
| $x_{ft}$ (mm) | 0.3453 | 0.3453 | 0.3363 |
| $T_{cv}$ (K) | 2763.4163 | 2763.4124 | 2767.7950 |
| $D_{CJ}$ (m/s) | 1977.1086 | 1977.0012 | 2050.8351 |
| $x_d$ (mm) | 0.1972 | 0.1972 | 0.1999 |
| Error | -- | $9.7508\times10^{-6}$ | $5.3725\times10^{-2}$ |
| CPU time (s) | | 113.32 | 21752.30 |

## 5. Discussion

The present ECA-NM method effectively leverages the global search capabilities of the ECA and the local optimization strength of the NM algorithm to calibrate the CDM parameters rapidly and accurately. This hybrid method demonstrates superior performance, simultaneously reducing the computational cost by approximately two orders of magnitude and diminishing the global error by nearly four orders of magnitude compared to the traditional GA-NM approach. This marked performance disparity is attributed to the intrinsic difference in the search mechanisms. The traditional GA depends on stochastic crossover and mutation operators to evolve the population. While effective for maintaining diversity, this stochastic nature lacks explicit directional guidance, often causing the population to oscillate around the optimum or become trapped in local minima. In contrast, the ECA employs an optimization strategy governed by the geometric center of mass. Instead of stochastic search, this mechanism exploits the spatial distribution of the population to construct a directional search vector. By synthesizing the coordinates of the superior individuals, the ECA calculates a center of mass that serves as a deterministic reference point, efficiently steering the search trajectory toward the most promising regions of the parameter space. This geometric guidance enables the algorithm to rapidly locate the vicinity of the global optimum, directly ensuring robust and precise convergence.

Although both methods incorporate the NM algorithm for local refinement, their overall efficiency is critically dependent on the quality of initialization provided by the global search phase. The NM algorithm, being a gradient-free direct search method, is highly sensitive to the initial guess and requires a starting point within the neighborhood of the global optimum. In this regard, the ECA yields a superior initialization characterized by significantly lower residual error compared to the GA. This high-quality starting point facilitates rapid local convergence for the NM algorithm, which directly explains the superior performance of the present ECA-NM method.

The superior convergence properties of the ECA-NM algorithm directly translate into the fidelity of the CDM across a broad operating range. As demonstrated in Section 4.2, the optimized parameters accurately reproduce the laminar flame speed and adiabatic flame temperature from fuel-lean to fuel-rich conditions ($\Phi$ = 0.5~2.5). This confirms that the method effectively captures the nonlinear dependence of flame and detonation properties on mixture composition, thereby extending the applicability of the CDM to non-uniform mixtures. Furthermore, the successful replication of the DTF evolution and the DDT in obstructed channels indicates that the calibrated reaction parameters remain valid under conditions of complex interactions among turbulence, shock waves, and chemical reactions. These results demonstrate that the proposed calibration method provides a robust parameter set, ensuring the predictive capability of the CDM for both deflagration and detonation regimes.

## 6. Conclusions

This paper presents and validates an efficient hybrid method combining the Evolutionary Center Algorithm (ECA) and the Nelder-Mead (NM) method to accurately calibrate the reaction parameters of the Chemical-Diffusive Models (CDMs). The ECA employs a center of mass strategy to efficiently find the vicinity of the global optimum, ensuring that the subsequent NM local search is initialized with a high-quality guess required for rapid convergence. Leveraging the global search capability of the ECA and the local optimization strength of the NM, the proposed method determines the optimal parameters of CDM that accurately reproduce the major properties of flame and detonation.

The CDMs for $H_2$-air and $H_2$-$O_2$ mixtures were developed using the hybrid method (ECA-NM) and validated against canonical tests of combustion waves and experimental data of flame acceleration (FA) and deflagration to detonation transition (DDT). Results show that the CDMs accurately reproduce their intended flame and detonation properties. Specifically, the laminar flame profiles calculated using the optimized parameters in 1D simulations exhibit excellent agreement with those calculated using a detailed mechanism over a wide range of equivalence ratios. The ECA-NM optimized parameters were then applied to 2D simulations, including distorted tulip flames, FA and DDT in obstructed channels. In these tests, the simulations accurately captured the complex flame morphological evolution of distorted tulip flames, and correctly reproduced the flame displacement history and DDT run-up distance as observed in experiments. These qualitative and quantitative agreements confirm the fidelity and reliability of the reaction parameters calibrated by the current method.

Finally, in terms of algorithmic performance, the ECA-NM method shows distinct advantages over the traditional GA-NM approach. By providing a high-quality initialization, the ECA enables the NM algorithm to achieve rapid local convergence, effectively avoiding the trap of local optima common in stochastic methods. Quantitatively, the ECA-NM method achieves significantly higher precision and efficiency of solution, reducing the global error by nearly four orders of magnitude and the computational cost by two orders of magnitude compared to the GA-NM method. This work provides a significantly efficient method for developing chemical-diffusive models that allows quantitative multi-scale simulations of flames, detonations, and transitions between them in complex scenarios.

**CrediT authorship contribution statement**

**Huahua Xiao:** Writing – review and editing, methodology, investigation, formal analysis, funding acquisition, supervision. **Xu Zhang:** Writing – original draft, Formal analysis, data curation. **Mingbin Zhao:** Writing – original draft, validation, conceptualization, methodology. **Congling Shi:** supervision, investigation.

## Declaration of competing interest

The authors declare that they have no known competing financial interests or personal relationships that could have appeared to influence the work reported in this paper.

## Acknowledgments

This work was supported by the Fundamental and Interdisciplinary Disciplines Breakthrough Plan of the Ministry of Education of China (Grant. No. JYB2025XDXM305) and Fundamental Research Funds for the Central Universities (Grant No. WK2320000061). We acknowledge the computing resources provided by the Supercomputing Center of University of Science and Technology of China (USTC).

## Supplementary material

Supplementary material is submitted with this manuscript. The supplementary file provides the optimal CDM parameters for $H_2$-air mixtures at equivalence ratios ranging from 0.5 to 2.5. It also includes a verification comparison of key flame and detonation properties obtained from the detailed mechanism and the calibrated CDM parameters.